%% file: main.tex
\documentclass[final,5p,times,twocolumn,authoryear]{elsarticle}

\usepackage{amssymb}

\usepackage{graphicx}
\usepackage{moreverb,url}
\usepackage{natbib}
\usepackage{xcolor}

\newcommand{\toolName}{{\textit{JobViz}}}

\journal{Visual Informatics}

\begin{document}

\begin{frontmatter}



\title{\toolName: Skill-driven Visual Exploration of Job Advertisements}

\author[label1,label2]{Ran Wang}

\author[label3]{Qianhe Chen}

\author[label5]{Yong Wang\corref{cor1}}
\cortext[cor1]{Corresponding author.}
\ead{yong-wang@ntu.edu.sg}

\author[label4]{Boyang Shen}

\author[label1]{Lewei Xiong}

\affiliation[label1]{organization={School of Journalism and Information Communication},
            addressline={Huazhong University of Science and Technology},
            city={Wuhan},
            % postcode={430074},
            % state={Hubei},
            country={China}}

\affiliation[label2]{organization={Philosophy and Social Science Laboratory of Big Data and National Communication Strategy},%Department and Organization
            addressline={Ministry of Education}, 
            city={Wuhan},
            % postcode={123}, 
            % state={123},
            country={China}}

\affiliation[label3]{organization={School of Computer Science and Technology},
            addressline={Huazhong University of Science and Technology},
            city={Wuhan},
            % postcode={430074},
            % state={Hubei},
            country={China}}

\affiliation[label4]{organization={Wuhan National Laboratory for Optoelectronics},
            % addressline={Huazhong University of Science and Technology},
            city={Wuhan},
            % postcode={430074},
            % state={Hubei},
            country={China}}

\affiliation[label5]{organization={College of Computing and Data Science},
            addressline={Nanyang Technological University},
            city={Singapore},
            % postcode={639798},
            % state={xxx},
            country={Singapore}}

\begin{abstract}
Online job advertisements on various job portals or websites have become the most popular way for people to find potential career opportunities nowadays. 
However, the majority of these job sites are limited to offering fundamental filters such as job titles, keywords, and compensation ranges. This often poses a challenge for job seekers in efficiently identifying relevant job advertisements that align with their unique skill sets amidst a vast sea of listings. 
% Guided by the feedback of in-depth interviews with job seekers, we propose a visual analytics approach, \toolName{}, to help job seekers efficiently explore the required skills and other relevant information of a large number of job posts. 
% Specifically, \toolName{} integrates visualization techniques with natural language processing (NLP) and other data processing techniques to facilitate interactive analysis and exploration of job posts. 
Thus, we propose well-coordinated visualizations to provide job seekers with three levels of details of job information: a skill-job overview visualizes skill sets, employment posts as well as relationships between them with a hierarchical visualization design; a post exploration view leverages an augmented radar-chart glyph to represent job posts and further facilitates users' swift comprehension of the pertinent skills necessitated by respective positions; a post detail view lists the specifics of selected job posts for profound   analysis and comparison.
By using a real-world recruitment advertisement dataset collected from 51Job, one of the largest job websites in China,
we conducted two case studies and user interviews to evaluate \toolName{}. The results demonstrated the usefulness and effectiveness of our approach.
\end{abstract}

\begin{keyword}
Visual Exploration \sep Job Advertisements \sep Skill-driven


\end{keyword}

\end{frontmatter}


% \section{}
% \label{}

\input{1-Introduction}

\input{2-RelatedWork}

\input{3-RequirementsAnlysis}

\input{4-DataAbstractionAndProcessing}

\input{5-SystemOverview}

\input{6-VisualDesign}

\input{7-CaseStudy}

\input{8-UserStudy}

\input{9-Discussion}

\input{10-ConclusionAndFutureWork}

\section*{Ethical approval}
Informed consent was obtained from all participants prior to their involvement in the study. Signed consent forms are on file and available upon request.

\section*{Acknowledgments}
This work was supported by the Huazhong University of Science and Technology teaching research project, Grant Number: 2023100, and the Fundamental Research Funds for the Central Universities, HUST: 82400049. The computation is completed in the HPC platform of Huazhong University of Science and Technology. Yong Wang is the corresponding author.


\bibliographystyle{elsarticle-num-names} 
\bibliography{template}





% \end{thebibliography}
\end{document}

%% file: 1-Introduction.tex
\section{Introduction}

Nowadays, companies are exploiting online technology (e.g., job portals and corporate websites) to make job advertisements reach ever-growing audience \citep{montuschi2013job}. Online advertisements on various job portals and websites have become the primary way for employers to reach 
% an ever-growing 
potential candidates and for job seekers to find suitable career opportunities~\citep{montuschi2013job}.
Job information generated on the Internet has shown an explosive growth \citep{si2021efficient}. 
However, it is difficult for job seekers to quickly understand the employers' demand and skill requirements for talents from the huge number of various job posts~\citep{si2021efficient}. Thus, how to deal with the recruitment information quickly and effectively is of great practical value for job hunting.

The most straightforward way is to use search engines to find information on the Web. However, it is time-consuming and inconvenient, especially for job seekers who need a clear job search scope \citep{zihan2021analysis}. They have to spend considerable time filtering job posts, as the results returned by search engines often contain a large number of job posts that users do not care about~\citep{ying2019data,si2021efficient}. Job seekers often adopt various special-purpose filters provided by recruitment websites (e.g., 51Job, indeed and LinkedIn)~\citep{si2021efficient}. These filters provided by recruitment websites can help job seekers improve their efficiency in processing job posts to some extent. However, after multiple rounds of in-depth interviews with three job applicants and two domain experts, we identified two major issues and concerns of hunting jobs on a recruitment website. On the one hand, though recruitment websites can provide preliminary filtering of job posts, it is still a challenge for job seekers to deal with a large number of roughly-filtered job posts, as they can hardly make a quick selection based on the overall properties of those job posts. To be specific, since all the filtered job posts are listed as separate items on different pages, job seekers have to check and compare key information one by one consequently instead of inspecting the similarities and differences of the filtered job posts as a whole.
On the other, it is necessary for job seekers to gain an in-depth understanding of the skills required and the relationships between different jobs. First, a job seeker concerns about not only his dream job but also similar jobs that he is qualified for in skills. Second, skills with a larger share of the market imply less risk of unemployment. Third, skill mismatch between job seekers and recruitment requirements may lead to employment difficulties~\citep{aasheim2012knowledge}. 
Thus, it is essential to find a proper job
% by means of 
via
skill exploration and matching, which, however, is difficult to achieve now. 
% \yong{Pls check my Chinese comments on overleaf.}

To address the above challenges, we propose a novel visual analytic system, \toolName{}, to achieve skill-driven visual exploration of job advertisements, helping job seekers analyze job posts efficiently and gain deep insights into the relationships between skills and jobs. In particular, the proposed skill framework in our paper can benefit bridging relationships between different job posts for further in-depth exploration.
By integrating natural language processing (NLP) and other data processing techniques with visualization techniques, we develop well-coordinated visualization views to provide users with the information of job posts at different levels of details: 
(1) a skill-job overview of job posts is displayed for filtering job posts rapidly, which visualizes skill sets, employment posts as well as the relationships between them with a hierarchical visualization design; 
(2) a post exploration view leverages an augmented radar-chart glyph to represent job posts and further enables users to gain a quick understanding of the corresponding job skills needed;
(3) a post detail view lists the details of selected job posts for further analysis and comparison.
Also, \toolName{} enables linked exploration and smooth interactions across the three major views so that our target users can explore the job posts from multiple perspectives and different levels of detail.

Taking the computer-science related jobs as an example,
we conducted case studies and user interviews to demonstrate the usefulness and effectiveness of our proposed \toolName{}, where real-world job posts 
% consisting of computer-science related recruitment advertisements 
collected from 51Job\footnote{\url{https://www.51job.com/}}, a popular online recruitment website in China, are used.
% \yong{Pls check if you are fine with it.}
The main contributions of this paper are summarized as follows:
\begin{itemize}
    \item An interactive visualization system, \toolName{}, to assist job seekers in analyzing and filtering recruitment information efficiently with multiple levels of details.
    \item A novel augmented radar-chart glyph to represent job posts in terms of both skills structure and posts distribution in a compact manner, facilitating interactive exploration and analysis of job posts. This innovative visual element enables users to quickly grasp the skill requirements and job posting patterns within specific job clusters, enhancing the efficiency of job market analysis and comparison.
    \item Case studies and user interviews with job seekers majored in computer science to demonstrate the usefulness and effectiveness of \toolName{} in helping users gain deep insights into skill-centered recruitment information online.
\end{itemize}

%% file: 2-RelatedWork.tex
\section{Related Work}

The related work of this paper can be categorized into two groups: data analytics of job advertisements and visualization of job posts.

% \textbf{Data Mining on Recruitment Websites:}
\textbf{Data Analytics of Job Advertisements:}
Recruitment advertising helps the employment market communicate its needs to job seekers~\citep{cullen2004lis}.
Thus, the analysis of job advertisements has become an important method for obtaining market demand except for questionnaires and semi-structured interviews. Existing research typically conducts content analysis of relevant advertisements, including industry type
% , job post
~\citep{debortoli2014comparing}, academic requirements~\citep{rios2020identifying}, knowledge and specific skills~\citep{suarta2018employability}, work experience~\citep{gibbons2020markers}, gender~\citep{gaucher2011evidence}, job category~\citep{gibbons2020markers}, country and region~\citep{debortoli2014comparing}, etc. While statistical data can provide an overview of job market demand, it is difficult to help job seekers find a specific job post matching their preferences from a large number of job posts.
% a huge amount of recruitment information.

% ``Skill" is a kind of important influence factors while looking for a job, and 
Skills
% are the key of any job post.
% Thus, they
are essential dimensions in the analysis of job advertisements. Employers often expect job seekers to have the necessary skills for the position at the start of their employment~\citep{abbasi2018analysis,zeidan2020effective,majid2019importance} and the ``skill gap" can affect one's ability to secure a job~\citep{radermacher2014investigating,abbasi2018analysis,aasheim2009knowledge,patacsil2017exploring}. Therefore, the emphasis on skills is both a necessary response to industry needs~\citep{pattanapairoj2021gap,baird2019employers,tan2018professional,zeidan2020effective} and an important guide to help job seekers identify effective learning paths in advance~\citep{rahmat2012relationship,zeidan2020effective}. However, when facing massive data, most researchers tend to reduce the volume of data or adopt word-based analysis, which can hardly extract systematical skill sets from recruitment information at the semantic level. In this paper, we propose a method based on NLP to extract multilevel and fine-grained skill sets from job advertisements effectively.

% \textbf{Visual Analytics of Recruitment Information:}
\textbf{Visualization of Job Posts:}
% There have been only a few works so far in this field. These existing 
Prior studies on the visualization of job posts can be roughly divided into two categories. One focuses on the visualization of overall recruitment information, including the geographical distribution of jobs ~\citep{si2021efficient}, salary levels in various regions~\citep{wan2020big}, majors and job matching status~\citep{2018Application}.
The common visualization techniques include maps~\citep{wan2020big}, tree diagrams~\citep{8457393} and radial bar chart~\citep{2018Application}. These approaches often ignore the skills required for the position and are too general to effectively guide job seekers to find matching jobs according to their abilities. The other is the visualization of skills required for specific positions, such as ship positions~\citep{logiodice2015spatial}, printing positions~\citep{zihan2021analysis}, and data analysts~\citep{ying2019data}, which are mainly displayed by word clouds~\citep{fang2021study, yang2019job}, histograms ~\citep{ying2019data}, and graphs~\citep{montuschi2013job}. These approaches often enable rich interactions, like highlighting, filtering, and floating panel, to show skills-related keywords. Nonetheless, they can offer an overview of job requirements, yet fail to provide an interactive way of dealing with recruitment information from the perspective of skills. Therefore, we propose a skill-driven visual exploration of job advertisements, facilitating job seekers to find job posts matched with their best in an interactive way.

%% file: 3-RequirementsAnlysis.tex
\section{Requirements Analysis} \label{Requirements}

To better understand the major challenges and design requirements for recruitment information analysis and job selection,
% and filtration, 
we have conducted several rounds of interviews with job applicants (P1-P3) and domain experts (E1-E2) to summarize their requirements in job hunting. 
% \yong{E1 and E2 cannot be categorized as ``job applicants''. 1) We may say ``with job applicants and domain experts''; 2) Make sure the abstract and introduction are synchronized.}
P1-P3 were three job seekers with different majors. P1 is a senior student majored in computer science and technology, wondering what kind of jobs he can be qualified for; P2 is a postgraduate majored in artificial intelligence, hunting a job matching his major; P3 is a senior engineer in a state-owned enterprise, looking for a new job in a different occupation. E1 is a director in charge of the Career Guidance and Service Center in a university for three years, and E2 is an assistant professor with research interest in higher education. Both of them are quite familiar with students hunting for jobs.
% \yong{E2 does not have any background relevant to job hunting.}
By conducting a series of interviews and discussions with them, we collected their feedback and summarized the major design requirements \textbf{R1-R4} as follows.

\textbf{R1. Explore the Overall Skill-centered Properties of Job Posts.} Based on the feedback of the five participants, all of them agreed that it is almost impossible to manually review all the recruitment information within limited time. According to their previous experience of job hunting, applicants with more matched skills with the occupation are of greater opportunities for the employment~\citep{cao2023technological}. Therefore, an effective approach should help easily and quickly inspect posts and corresponding skills structure required. In particular, E1 pointed out that it was critical to understand the job market and its required skills as a whole for one's career development planning. P2 suggested that some properties of job posts, such as salary, location, qualification, experience and industry, are supposed to be considered for finding appropriate jobs.  

\textbf{R2. Inspect Skill Patterns and Relations of Different Job Posts.} All the particiapnts agreed that the required skills of job posts can provide significant clues for job selection. According to E2's experience, versatile talents with multiple skills were more preferred in the job market. For example, equipped with several professional skills, a undergraduate student majored in journalism could be competent for marketing, advertisement, publishing, operation position, product manager, public relations and so on. P1 pointed out that he tends to choose a job that matches his skills best, especially for circuits and electronics that he is good at. P2 and P3 also  emphasized the need for a drill-down exploration of skill patterns needed by the filtered job posts, because skill mismatch may lead to negative impact on one's career according to his previous internship experience. In addition, they also took the skills that were popular in the job market into consideration, since the popular job skills meant more job opportunities and less risk of unemployment in the future~\citep{deming2018skill}.
Therefore, comparing different job posts’ skills and inspecting their similarity and difference are very valuable for job seekers. 

\textbf{R3. Compare Key Properties of Job Posts with Similar Skill Patterns.} P2 and E2 pointed out that individual job posts with similar skill patterns were supposed to be inspected in depth. In particular, those jobs highly relevant to the specific skills that one was good at were more preferred, and the selection process could benefit from the distribution of job posts regarding skills, because people tended to foster strengths and circumvent weaknesses. P1 and P3 highlighted that job seekers also consider other factors beyond skill requirements, such as location, work experience, salary and industry domain.
% work experience, 
% company type, business sector, job responsibilities, and employee benefits. 
This aligns with previous research findings~\citep{rafaeli2005recruiting} on job seekers' information needs. For example, P3 wants to change to another job in the same city yet in a different field, and a job in a private company with higher salary while maintaining a similar skill structure is more appreciated by him. Thus, visual comparison of filtered job posts' key properties can help job seekers select proper jobs.

% it is also important to be informed of key properties for job posts.
% \textbf{R2. Explore job posts Through Characteristics Comparisons.} All the interviewees agreed that the characteristics analysis of job posts can provide significant clues for job selection. E1 pointed out that he tends to choose a job that matches his skills best, especially for circuits and electronics that he is good at. In addition, E3 wants to change to another job in a different field, and a job with higher salary while maintaining a similar skills structure is more appreciated by him. Thus, visual exploration of filtered job posts sorted by different characteristics can help job seekers select proper jobs in candidate.

\textbf{R4. Show the Details of Individual Job post.} As the five participants suggested,
% a functionality of display and compare the details of selected job posts is essential. 
it is essential to explore and compare the details of selected job posts~\citep{aasheim2009knowledge}.
It can help job seekers further analyze and confirm their preferred jobs through listed recruitment information in detail, such as location, company category, welfare, etc.

%% file: 4-DataAbstractionAndProcessing.tex
\section{Data Abstraction and Processing} \label{data}
The job post dataset used in this paper is collected from 51Job, one of the most influential human resource service providers in China~\citep{2022HRoot}.
As a proof of concept, we focus on the job posts related to computer science and engineering.
Retrieving with the keyword ``computer'', we collected about 2.43 million job posts in total from July 2019 to July 2021.
% from 51Job. 
A job post includes post information, salary, location, industry, company information, etc.

\begin{figure*}[ht]
    \centering
    \includegraphics[width=\linewidth]{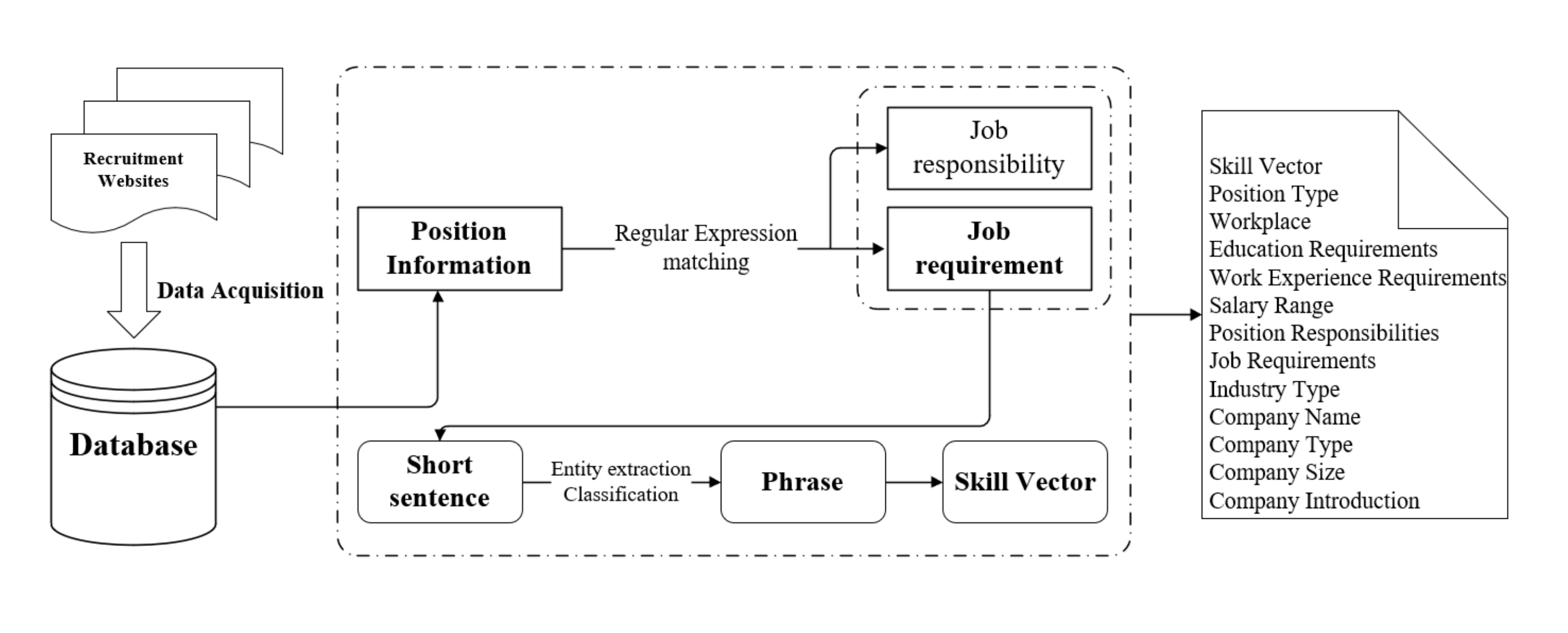}
    \caption{The technical framework for skill extraction.}
    \label{fig:liuchengtu}
\end{figure*}

\textbf{Skill Framework.} According to the existing competency model~\citep{2020Computing,2020Operationalisation}, the skill framework of computer science and engineering
% can be divided into three hierarchies. 
consists of three levels of hierarchies.
The first level contains technical and foundational skills. In the second level, technical skills contain Users and Organizations, Systems Modeling, Systems Architecture and Infrastructure, Software Development, Software Fundamentals and Hardware. Additionally, foundational skills are classified as Personal and Interactive ones.
The third level of the hierarchy contains 15 detailed technical skills, which, according to common skill taxonomy~\citep{2020Computing,2008Critical,2008Mapping},
% On the basis of common skill classification framework~\cite{2020Computing,2008Critical,2008Mapping}, 20 technical skills in the third hierarchy
can be identified as Social Issues and Professional Practice, Data and Information Management, Intelligent Systems (AI), Software Design, Operating Systems, Architecture and Organization etc. Likewise, 8 foundational skills are defined as: Personal skill, Mathematics and Statistics, Enterprising Skill, Communication and Presentation, Leadership Skill, Interpersonal Skill, Organisational Skill and Team Skill~\citep{2018Identifying,radermacher2014investigating,aasheim2009knowledge,jotse271}.

% On the basis of CC2020\cite{2020Computing} and the skill classification framework\cite{2008Critical}\cite{2008Mapping} derived from the skill needs of the market recruitment with the common skill categories in the recruitment advertisement, the 20 technical skills are as follows: Social Issues and Professional Practice, Data and Information Management, Intelligent Systems(AI), Software Design, Operating Systems, Architecture and Organization etc.

% Foundational skills are introduced based on the skill requirement framework for graduate recruitment of IS (Information System) majors\cite{2018Identifying}.
% Among the non-computer professional skills, the skills of using enterprise management systems such as ``ERP", ``SAP" and ``OMS" frequently appear in the skills of market recruitment data, so the ability of ``Organisational Skill" is increased.
% The work communication and expression skills of graduates (such as oral communication, document writing, presentation etc.)\cite{2014Investigating}\cite{2009Knowledge}\cite{jotse271} when they first enter the job often do not meet expectations, so ``Communication and Presentation" are added to it. According to the curriculum and the market, entrepreneurial ability is involved, and ``Enterprising Skill" is added to soft skills. To sum up, the 8 foundational skills are as follows: Mathematics and Statics, Mathematics and Statics, Enterprising Skill, Communication and Presentation, Leadership Skill, Interpersonal Skill, Organisational Skill and Team Skill.

\textbf{Skill Extraction.}
To extract skills in demand from a huge amount of job advertisements, we propose to first identify the effective information from job advertisements for key phrases, which are then characterized as the corresponding skill categories. The technical framework for skill extraction is shown in Figure~\ref{fig:liuchengtu}.Regular Expression Matching is used to filter job requirements. Then, to classify millions of short sentences in job requirements into three categories, i.e., technical skill, foundational skill and irrelevant information, a set of 3000 class-balanced labeled data was randomly selected through manual annotation firstly, and subsequently utilized to train network classifier.
Our system involves analyzing a dataset comprising 12 million entries (representing 2.43 million job listings, each with an average of 6 entries, where each entry corresponds to a specific job requirement). Given this considerable volume, the BERT\citep{kenton2019bert} model proves more advantageous than LLMs in terms of reduced processing times and lower costs. We anticipate incorporating more cutting-edge feature extraction techniques in future iterations. Subsequently, a softmax classifier is deployed to conduct a three-class classification. After training, the classification attained an accuracy of 93.7\%. Taking into account both the intrinsic errors in the original data due to input mistakes and the favorable feedback obtained from participants and experts during the system's pilot phase, this accuracy level is deemed satisfactory.
%The model employed a pretrainedBERT~\citep{kenton2019bert} for text feature extraction and implemented a softmax classifier for three-class classification. Following the training process, a classification accuracy of 93.7 was achieved. 
Consequently, the trained classifier was then applied to classify the remaining unlabeled data.

Next, we employed the entity extraction method~\citep{2010Automatic,2017Salience} provided by the JioNLP, a Chinese NLP prepossessing tool, to extract key phrases from short sentences in job requirements. Since the skill classification, which is based on Chinese words or phrases obtained in the previous step, is too complicated for machines to guarantee high accuracy, manual classification is adopted for this classification task. Thus,
% three trained people participate in classifying 5,000 key phrases jointly, 
we first recruited three well-trained research assistants to classify 5,000 key phrases independently. If the classification labels for a key phrase by all of them are consistent, the label is accepted. For the remaining key phrases without a consistent label, we further employed a domain expert (E2) to review and decide the final labeling results to ensure the classification accuracy.
% we recruited three well-trained research assistants to classify 5,000 key phrases jointly, 
% with a reliability test that can prove the result is credible.
Therefore, a dictionary of skills can be constructed, containing classified key phrases for skills in accordance with the framework mentioned above. It should be noted that when the same skill appears multiple times in each job requirement separated by a serial number, the degree of skill demand is calculated only once. Finally, a normalized skill vector for each job advertisement can be obtained for further analysis, in which each number refers to the corresponding skill proportion.
%\wy{I do not understand the revision.}
The similarity between two job posts can be calculated based on the Euclidean distance of both skill vectors.
% In particular, to gain the skill proportion, we also calculated the proportion of different skills in each job post.

\textbf{Job Clustering.}
To reveal the patterns of skills required by one kind of job post, an unsupervised clustering method, namely, affinity propagation~\citep{AffinityPropagation}, was employed to divide the selected job advertisements by users into different clusters based on their corresponding skill vectors. The primary advantage of selecting the affinity propagation over other clustering methods like K-means and DBSCAN is that it eliminates the need to predefine the number of clusters, which is particularly beneficial when dealing with recruitment data where predicting both quantity can be challenging. By automatically determining the optimal number of clusters, the affinity propagation clustering method allows for a more flexible approach to processing and comprehending data.

%% file: 5-SystemOverview.tex
\section{System Overview}

\begin{figure*}[ht]
  \centering
  \includegraphics[width=\linewidth]{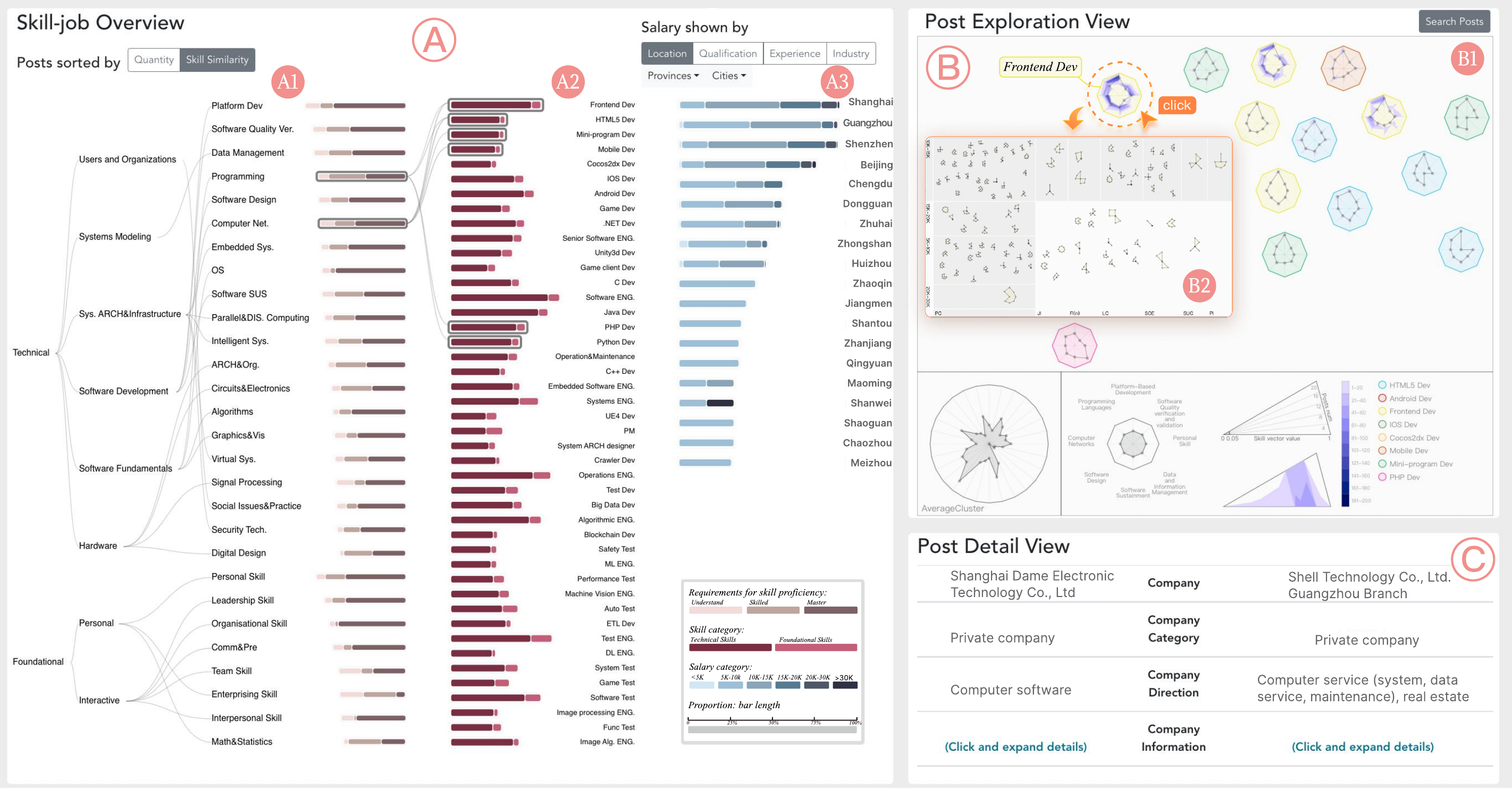}
  \caption{\toolName{}, a skill-driven visual analytics system to help job seekers to efficiently explore the required skills and other relevant information of a large number of job posts in an interactive way. A) a skill-job overview of job postings is displayed for filtering job postings rapidly, which visualizes skill sets, employment posts as well as relationships between them; B) a post exploration view leverages a metaphor-based glyph to represent each job post and further enable users to gain a quick understanding of their key properties; C) a post detail view lists the details of selected job posts for further analysis and comparison.}
  \label{fig:toutu}
\end{figure*}

\begin{figure}[ht]
    \centering
    \includegraphics[width=\linewidth]{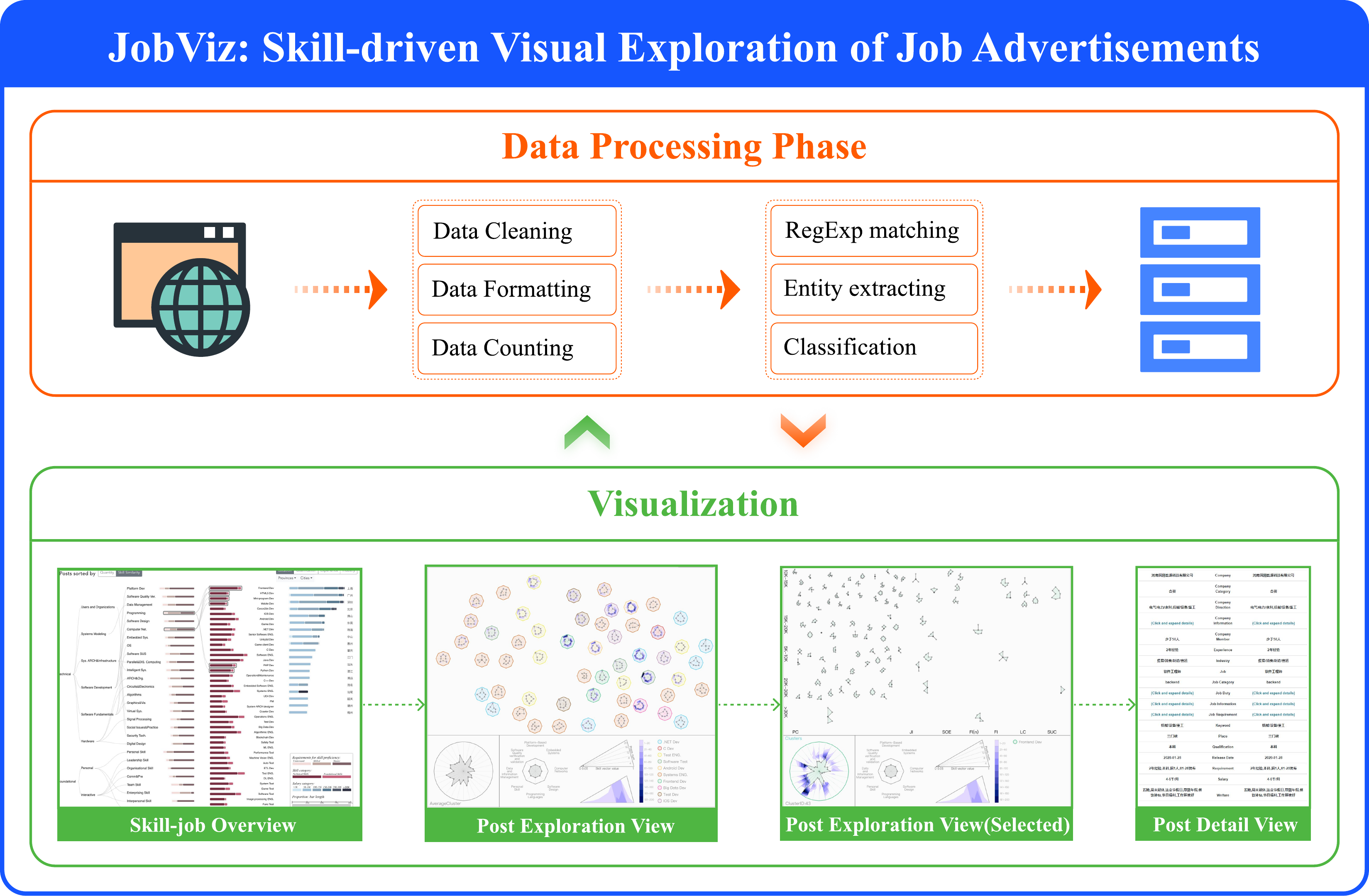}
    \caption{The system architecture of \toolName{} contains three modules: a storage module, a processing module, and a visualization module.}
    \label{sysstructure}
\end{figure}

Figure~\ref{sysstructure} shows the pipeline of our system, which consists of two main parts, i.e., data processing and visual exploration. The first phase processes millions of original job advertisements and extracts required skills and key properties using NLP algorithms. The details are provided in Section \ref{data}. Then, based on the four major design requirements (Section \ref{Requirements}), we design an interactive visualization system, which can help job seekers to efficiently explore the required skills and other relevant information of a large number of job posts.

We implemented a web-based system based on the Vue.js front-end framework and the Flask back-end framework. Figure~\ref{sysstructure} shows the architecture of \toolName{}, which consists of three major views, namely, skill-job view (Figure~\ref{fig:toutu}A), post exploration view (Figure~\ref{fig:toutu}B1-B2) and detail view (Figure~\ref{fig:toutu}C). The skill-job view provides a visual summary of all the job advertisement of computer science, allowing users to explore the relationship between skill sets and job information (\textbf{R1}). The post exploration view comprises two parts: job cluster and post map. The job cluster supports the inspection of skill patterns of different job clusters (\textbf{R2}), including skill structures and post distribution for each skill. Also, users can further visually check and compare individual job posts within a cluster in terms of corresponding key properties in the post map (\textbf{R3}). Post Detail View provides a table, listing detailed information in the selected job post (\textbf{R4}).

%% file: 6-VisualDesign.tex
\section{Visual Design}

% Figure~\ref{sysstructure} demonstrates the whole pipeline of our system including the data processing phase and the visual exploration phase. The first phase processes millions of original job advertisements and extracts required skills and key properties using NLP algorithms. The details are provided in Section \ref{data}.

% We propose \toolName{}, an interactive visualization approach to to help job seekers to efficiently explore the required skills and other relevant information of a large number of job posts.

% \begin{figure}[ht]
%     \centering
%     \includegraphics[width=0.75\textwidth]{figures/sysstructure.jpg}
%     \caption{The system architecture of \toolName{} contains three modules: a storage module, a processing module, and an visualization module.}
%     \label{sysstructure}
% \end{figure}

% Figure~\ref{sysstructure} shows the architecture of \toolName{}, which consists of three modules: storage module, processing module and visualization module. The storage module stores the original job advertisements collected from 51Job, including post information, salary, location, industry, company information, etc. The processing module handles NLP algorithms to extract skill and other key properties from those job advertisements for exploration and comparison. The visualization module reveals the relationship between job posts and skills needed.

% As mentioned above, especially according to the feedback from P1-P3 and E1-E2, 
Built upon the above design requirements, we propose intuitive visual designs to help job seekers analyze and filter recruitment advertisements based on job posts, required skills, and corresponding properties. 

\subsection{Skill-job Overview}

% Skill-Job Overview (Figure~\ref{fig:toutu}(A)) is designed to provide job seekers with an overview of all job postings in terms of skill-job distribution and relationships between them \textbf{(R1)}. It is of great significance since multiple properties of job posts as well as their relationships can be displayed and explored for matching and filtering job posts. 

Skill-job Overview (Figure~\ref{fig:toutu}(A)) hierarchically represents a visual summary of all the posts in job market of computer science, and allows the user to explore from three aspects: skill sets, job posts and properties concerned (\textbf{R1}). It is of great significance since skills and related posts can be displayed and explored for job seekers to match and filter jobs. 
% We propose a novel parallel coordinates chart
Inspired by high-dimensional data visualization~\citep{2015Interactive} and spatio-temporal data visualization~\citep{7192676}, we propose a hierarchical visual design based on parallel coordinates chart and tree diagram
and it consists of three parts in tight connections: a design of a tree and bar chart to visualize skills framework and requirements jointly (Figure~\ref{fig:toutu}(A1)), and two stacked horizontal bar charts to display different job posts and corresponding properties respectively (Figure~\ref{fig:toutu}(A2 and A3)). The interactions between these views are performed in a coordinated way.

% \begin{figure}[ht]
%     \centering
%     \includegraphics[width=0.35\textwidth]{figures/OverviewLegend.pdf}
%     \caption{Work process and Technical route}
%     \label{legend}
% \end{figure}

The left part contains a tree and a stacked horizontal bar chart, as shown in Figure~\ref{fig:toutu}(A1). The former shows extracted skill sets at three hierarchies, and the latter displays the corresponding skill share in market and proficiency. In particular, each stacked bar encodes two different metadata, length for skill share in market and color for its proficiency, shown in Figure~\ref{fig:toutu}. The stacked bars are vertically ordered according to their proportions, starting from the top, allowing the job seeker to see the whole skill structure and requirement at a glance. Within each bar, it is further divided into three sections from left to right, representing ``understand'', ``skilled'', and ``master'' levels of proficiency respectively. The middle part shown in Figure~\ref{fig:toutu}(A2) represents posts in the job market of computer science, in which the length of a bar encodes the market share of the job post. It should be noted that the bar contains two parts horizontally, with technical skill on the left and fundamental skill on the right. Likewise, job posts are also vertically ordered according to their market share by default. Thus one could easily notice the differences among job posts in terms of quantity and skill. The right part shown in Figure~\ref{fig:toutu}(A3) displays location, qualification, experience and industry distribution related to the filtered job posts, allowing the job seeker to further filter job posts under different conditions (\textbf{R4}). The length of the bar encodes how many job posts under the selected constrains. Four distinctive qualitative colors as well as the length of stacked blocks in horizon within a bar, are applied to encode salary categories and their proportions respectively. The bar shown in Figure~\ref{fig:toutu}(A3) displays the salary distribution for different categories of a selected attribute (e.g., experience levels, qualifications, locations, or industries). The gray-to-blue color scheme represents different salary ranges, with darker blue indicating higher salaries. Users can switch between attributes using the dropdown menu at the 
% top of the chart 
topright corner of Figure~\ref{fig:toutu}(A3)
to explore salary trends across different job aspects. By the way, the payment can note be encoded as a linear scale because most job posts don't provide the salary in specific. Overall, job posts and their required skills and properties distributions are connected via curve links, creating a systematic view as a whole. 

Corresponding interactions are supported in Skill-job Overview. First, a panel would appear to display a brief word cloud that can represent the skill when hovering on the corresponding skill, which enables the job seeker to learn more about the skill in a simple but effective way. Second, the relationships between job posts and skills are revealed though links in gray color, shown between Figure~\ref{fig:toutu}(A1) and Figure~\ref{fig:toutu}(A2). If a job post is selected, and then all required skills of it will be highlighted and connected to it. Meanwhile, if a skill is selected, all job posts that need this skill will be highlighted and connected to it in the same way. Third, job posts can be reordered in the vertical direction based on skills similarity, which could be calculated by the Euclidean distance between two skill vectors. For example, after right clicking on ``Python Dev'', it would be displayed on the top as the first item, followed by ``PHP Dev'', ``Java Dev'', ``Crawler Dev'', ``C Dev'', so on and so forth. Those job posts with a more similar skill structure with ``Python Dev'' tend to be closer to it in spatial position. Thus, the job seeker can be provided with more career options feasibly based on his skill structure. Fourth, the job seeker is able to filter certain posts according to location, qualification, experience and industry. By means of clicking on the filter in Figure~\ref{fig:toutu}(A3), all job posts he is interested will be fed into the next Post Exploration View for further inspection.

% \textit{Design Alternatives:} 
Before finalizing the current visual design, we also considered other alternative designs like parallel coordinates plot \cite{hewes1883scribner} and tree diagram \cite{SurveyMultipleTree}. In our scenario, each skill in the framework may be simultaneously required by several job posts in the market, which need a set of skills in turn as well. However, in a parallel coordinates plot, each instance is represented by a line that connects a point in each parallel axis, which makes it inapplicable to display multiple relationships between skills and job posts. In addition, it is a many-to-many relationship among the skills, job posts and their properties, which is not supported by a tree diagram.
% \yong{..... Please list concrete reasons on why the PCP and TreePlus graph cannot be used for effectively providing an overview of all the job posts. }
% Alternative designs of the factor view have been considered. The parallel coordinates chart and TreePlus graph~\cite{2006Treeplus} are the most intuitive forms to visualize skills, job posts and relationships between them. And our current design differs from two aspects. On one hand, the items within can be rearranged in accordance with specific conditions, i.e., quantity or skill similarity. On the other, one item can be connected to many items, which is very different from the parallel coordinates chart. 
Overall, our current design can facilitate exploring and filtering skills and job posts efficiently and flexibly.

\subsection{Post Exploration View}
Post Exploration View (Figure~\ref{fig:toutu}(B)) is designed to help job seekers gain a quick understanding of key properties belonged to the filtered job posts for in-depth comparison and selection (\textbf{R2 and R3}). This view comprises two main components: the job cluster view (Figure~\ref{fig:toutu}(B1)) and the post map view (Figure~\ref{fig:toutu}(B2)), which are displayed sequentially. Users first interact with the job cluster view to select a cluster of their interest. Upon selection, the entire job cluster view (B1) is replaced by the post map view (B2), allowing users to further explore and interact with the selected cluster in more detail.
The post exploration view consists of glyphs representing the filtered job posts projected onto a 2D skill map, where the distance between glyphs indicates the similarity of skill structures needed by different job posts. Especially, we propose a novel augmented radar-chart glyph to represent each job post cluster in terms of both skills structure and posts distribution in a compact manner, facilitating interactive inspection and comparison of job posts.

\begin{figure*}[htbp]
    \centering
    \includegraphics[width=\linewidth]{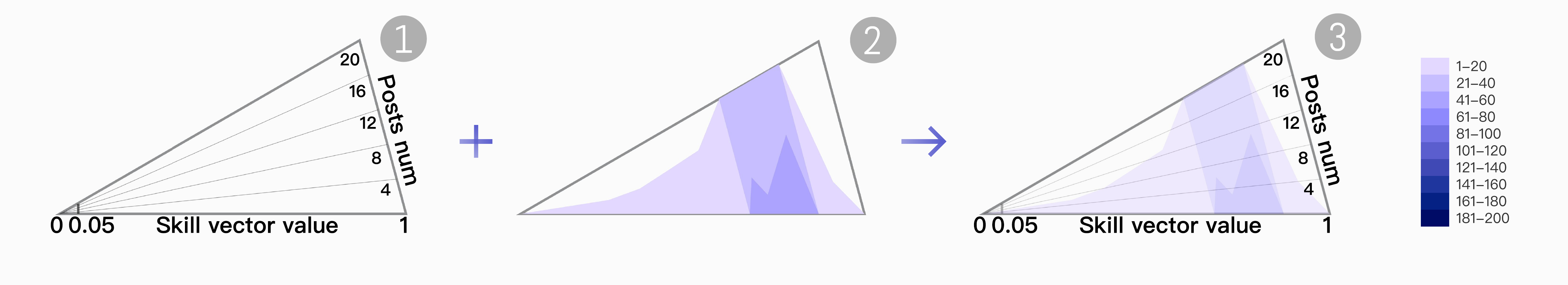}
    % \caption{Illustration for how to combine a horizon chart with a sector in a radar-chart to show the job posts distribution in a compact manner.}
    \caption{An illustration of our augmented radar-chart glyph to show individual job posts in a compact manner, where the glyph integrates the horizon chart design. Sub-figure1 indicates a sector of a radar chart with spokes representing specific skills; Sub-figure2 indicates a horizon chart containing job posts distribution within; Sub-figure3 indicates a combination of them.
    % \yong{It may be better to explain Subfigures 1, 2 and 3 as well.}
    }
    \label{legend}
\end{figure*}

\textbf{Job Cluster}: We first cluster all the filtered job posts based on their skill vectors, and each cluster is encoded as an augmented radar-chart with a specific color, which consists of a radar chart and several horizon charts, as shown in Figure~\ref{fig:toutu}(B1). In particular, the radar chart includes a sequence of equi-angular spokes, with each spoke representing one specific skill. The length of a spoke is proportional to the magnitude of the average proportion of each skill in a cluster, and a line is drawn connecting the data values for each spoke. It should be noted that only representative skills instead of all of them are shown for space saving, yet they still add up to more than 80\% in total. In addition, we visually summarized the overall distribution of job posts for each skill in a cluster inspired by the horizon chart, as shown in Figure~\ref{legend}. In each sector in a radar chart, an area chart is sliced into equal intervals along the tangent direction of the sector and collapsed down into single bands, which makes the glyph more compact and similar to a heat-map. Especially, all the bands beyond the baseline along the baseline are flipped and then stacked on top of each other. The darker colors indicate more job posts corresponding to this skill ratio. Thus, with combination of the radar chart and horizon chart, job posts in terms of both skills structure and posts distribution can be displayed in a compact manner.

\textbf{Post Map}: During several rounds of interviews with applicants, they mentioned that some key properties of job posts can also provide important reference for job hunting, such as salary and company category. To this end, by clicking one cluster in Figure~\ref{fig:toutu}(B1), we locate all the job posts in the cluster on a 2D map for further inspection according to their key properties, i.e., company category in horizon and salary in vertical, as shown in Figure~\ref{fig:toutu}(B2). Vertically, they are distributed according to the quantity of job posts with the same salary. It worth noting that those job posts sharing the same properties are located randomly within an area determined by the both properties. 
To address scalability concerns, especially when dealing with a large number of job postings, our approach leverages a concise glyph system design. The key principle is that users can effectively compare and evaluate options by assessing the overall morphology of the glyphs, rather than scrutinizing their intricate details. This design transforms the glyphs in Figure~\ref{fig:toutu}(B2) 
into rapid visual indicators of job characteristics, enabling users to swiftly identify and compare suitable positions even when there are multiple matched job positions.
% presented with a multitude of potential matches.

In particular, a radar chart is employed to encode each job post, keeping its shape and color in accordance with glyphs in Figure~\ref{legend}. If the vertex of a spoke is close to the center of the radar chart, it means that the corresponding skill proportion is close to 0. By this way, Post Exploration View enable job seekers gain a quick understanding of their concerned job posts in terms of key properties, facilitating further comparison and selection.

\subsection{Post Detail View}

Post Detail View (Figure~\ref{fig:toutu}(C)) helps job seekers inspect details of selected posts for decision making (\textbf{R4}). It consists of a table, listing detailed recruitment information such as location, company category, job duties, welfare, etc. Meanwhile, two job posts can be compared in detail simultaneously by clicking two candidate posts in Post Exploration View. With combining all the information, several posts that best matched with the job seeker could be selected.
% \yong{This subsection may be too short to be a subsection.}

%% file: 7-CaseStudy.tex
\section{Case Study}

We conducted two case studies on an online recruitment dataset collected from 51Job, to demonstrate the effectiveness of \toolName{}. The users involved in the case studies are two participants (P4 and P5).
% mentioned in Section~\ref{Requirements}. 

\subsection{Case Study 1 - A Senior Student}

P4 is a senior student and
% , a senior student majored in computer science and technology, 
will graduate soon. He is actively seeking a desirable job and we invited him to use \toolName{} to find a suitable job advertisement matching his background and job preference. % that matched with him. 

\textbf{Selecting job posts based on skills similarities and other properties as an overall consideration.} 
The first goal of P4 was to observe the overall distribution of jobs, skills and relationships between them. Such kind of information can help him know what jobs he might be qualified in the job market of computer science. Thus, P4 initiated interaction with the Skill-job Overview  in Figure~\ref{fig:toutu}(A). He found the skill tree has been well established and displayed from ``hard skill'' to ``soft skill'', as shown  in Figure~\ref{fig:toutu}(A1). After keeping browsing each skill by hovering on, the corresponding word clouds were shown one by one, helping him learn more about meanings of these skills briefly and rapidly. He found that many of them have been taught in his college time, which made him feel a little bit confident. In addition, traditional skills were usually of darker color, which meant higher proficiency was required, compared with emerging ones. \textbf{Next, he clicked a skill, namely ``Programming and Computer Net'', and found that all job posts needing this skill were highly associated  with computer network software development, which were highlighted in Figure~\ref{fig:toutu}(A2).} Afterwards, P4 right clicked the post ``Frontend Dev'' he was interested in, and wanted to know what posts else required similar skills. The result was exactly as he thought, and the top rankings were all job posts related to web or mobile application development. Additionally, salary was another important issue he concerned. Unlike social recruitment, which mainly required work experience, campus recruitment paid more attention to academic qualifications. Therefore, P4 first chose to display salary distribution in accordance with academic qualifications, as shown in Figure~\ref{fig:toutu}(A3). It was found that the average salary of one with bachelor degree provided by the job posts he selected was around RMB 10000-15000 per month. Subsequently, he chose to display the salary distribution in different cities, like Beijing, Shanghai, Shenzhen, etc., where more work opportunities with higher salary were provided. By observing the color blocks in the middle occupy the main space, it could be deduced that salaries in these cities were basically above 10k, which satisfied his psychological expectations. Therefore, P4 wanted to continue to explore job posts in individual on Post Exploration View.

\begin{figure}[ht]
    \centering
    \includegraphics[width=\linewidth]{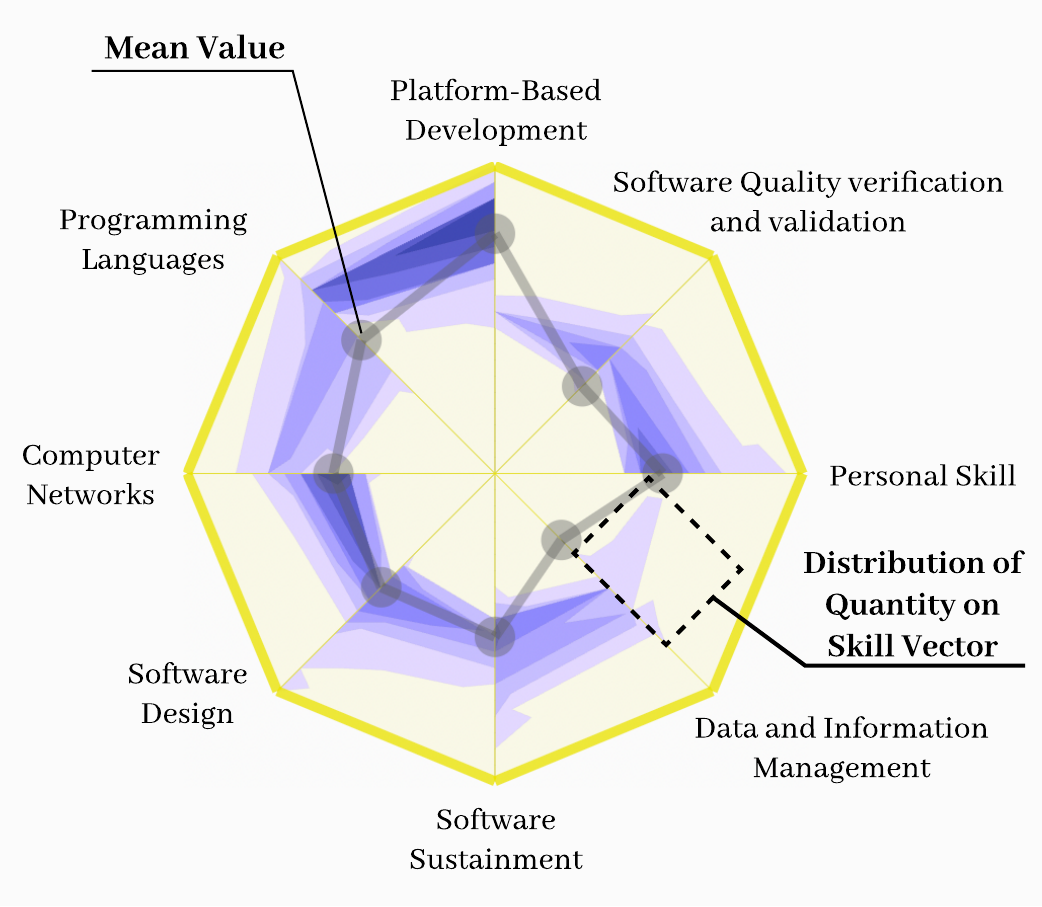}
    \caption{A new glyph design augmented from radar-chart and horizon chart is proposed to represent each job post cluster in terms of both skill structures and post distributions in a compact manner. 
    % \yong{pls move the figure close to the corresponding text description}
    }
    \label{case11}
\end{figure}

\begin{figure}[ht]
    \centering
    \includegraphics[width=\linewidth]{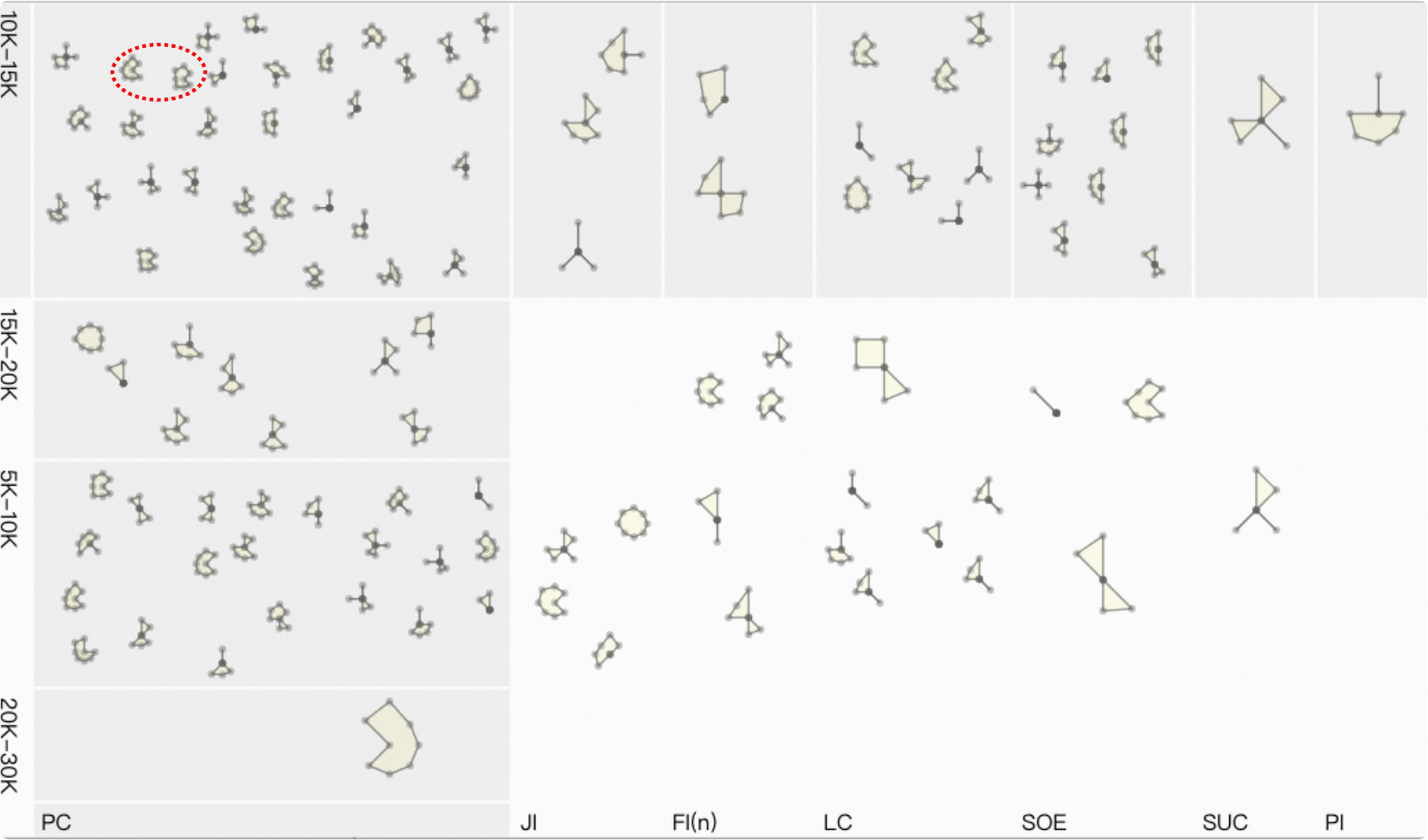}
    \caption{The Post Map in Case Study 1. Two job posts that matched P4's skills were chosen. 
    % \yong{pls move the figure close to the corresponding text description}
    }
    \label{case1post2}
\end{figure}

% \textbf{Filtering job posts based on skill distribution and salary level.} 
\textbf{Choosing promising positions based on skill distribution and salary level.} 
After selecting the candidate job posts, then P4 starts interacting with Post Exploration View to inspect specific job posts for further filtering. At the beginning, he gained a quick understanding of the legend to figure out each skill in a radar-chart. Then P4 excluded those clusters with few job posts, which means less opportunities to be offered, and focus on clusters with relatively dense job posts for further exploration. For a cluster of ``Frontend Dev'', he found that most of job posts were distributed around mean values, i.e., those job posts were quite similar in terms of skills, as shown in Figure~\ref{case11}. In his opinion, he can obtain more opportunities in this field and clicked the cluster into post map to inspect more details. P4 noticed that each specific job post was distributed on the post map according to the salary level and company category, as shown in Figure~\ref{case1post2}. He planed to work in a private company for a higher salary level, he checked all the clusters with 10000-15000 payment in the block of ``PC''. The salary offered by these jobs might not be the highest, but they did provide the largest pool employment opportunities, thereby offering a wider range of choices. And then he chose two job posts that matched his skills for comparison. Combining all of the above information, P4 decided to move to Post Details View for gaining more recruitment information of both job posts for final decision.

% \textbf{Analyzing the details for final decision.}
\textbf{Detailed Comparison for final-decision making of job application.}
In the Post Detail View, P4 compared the two job posts selected from the Post Exploration View, as shown in Figure~\ref{fig:toutu}(C). He found that both job posts are highly related to his major and he could be qualified with the skills required. Besides, by considering the proper city and excellent welfare, he believed that both jobs satisfied his expectation and were worth submitting his application.
% sending his resume.

\subsection{Case Study 2 - A Senior Engineer}

P5 is a senior engineer who just quit his job as an Android developer
% with seven years experience 
at an Internet company, and is looking for a leadership post and a better working environment with \toolName{}. 
% He was invited to use \toolName{} to search a suitable job post. 

\begin{figure}[ht]
    \centering
    \includegraphics[width=\linewidth]{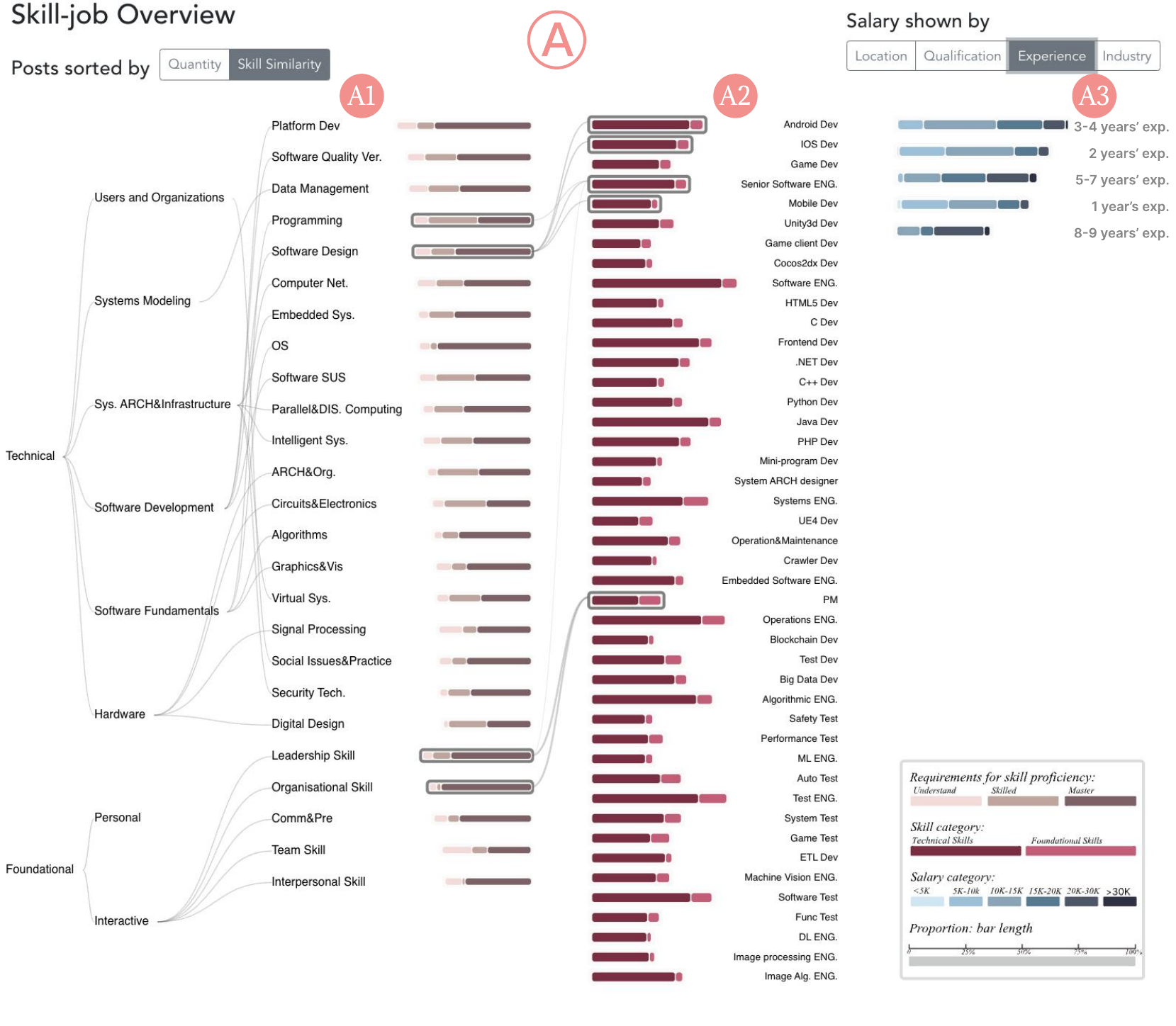}
    \caption{The Skill-job Overview in Case Study 2, in which P5 filtered job posts according to his matched skills and working experience.}
    \label{case2posts}
\end{figure}

% \textbf{Selecting job posts with a clear purpose.} 
\textbf{Selecting job posts with a clear purpose from a technical position to a managerial one.} 
As shown in Figure~\ref{case2posts}, P5 first clicked three technical skills like ``Programming'', ``Software Design'' and ``Computer Net'' which he was quite familiar with. Then, ``Leadership Skill'' and ``Organisational Skill'' were also selected because both skills would be required by the jobs he was looking for. Due to the extensive experience in Android development, P5 preferred to work in his original field, and hence clicked the job post ``Android Dev'' and some similar posts (i.e., ``IOS Dev'', ``Senior Software ENG'', ``Mobile Dev'', ``PM''). As a senior engineer, P5 was not sensitive to the factors of education, location and industry, and thus only selected the job posts in need of 5-7 and 8-9 years of working experience.

\begin{figure}[ht]
    \centering
    \includegraphics[width=\linewidth]{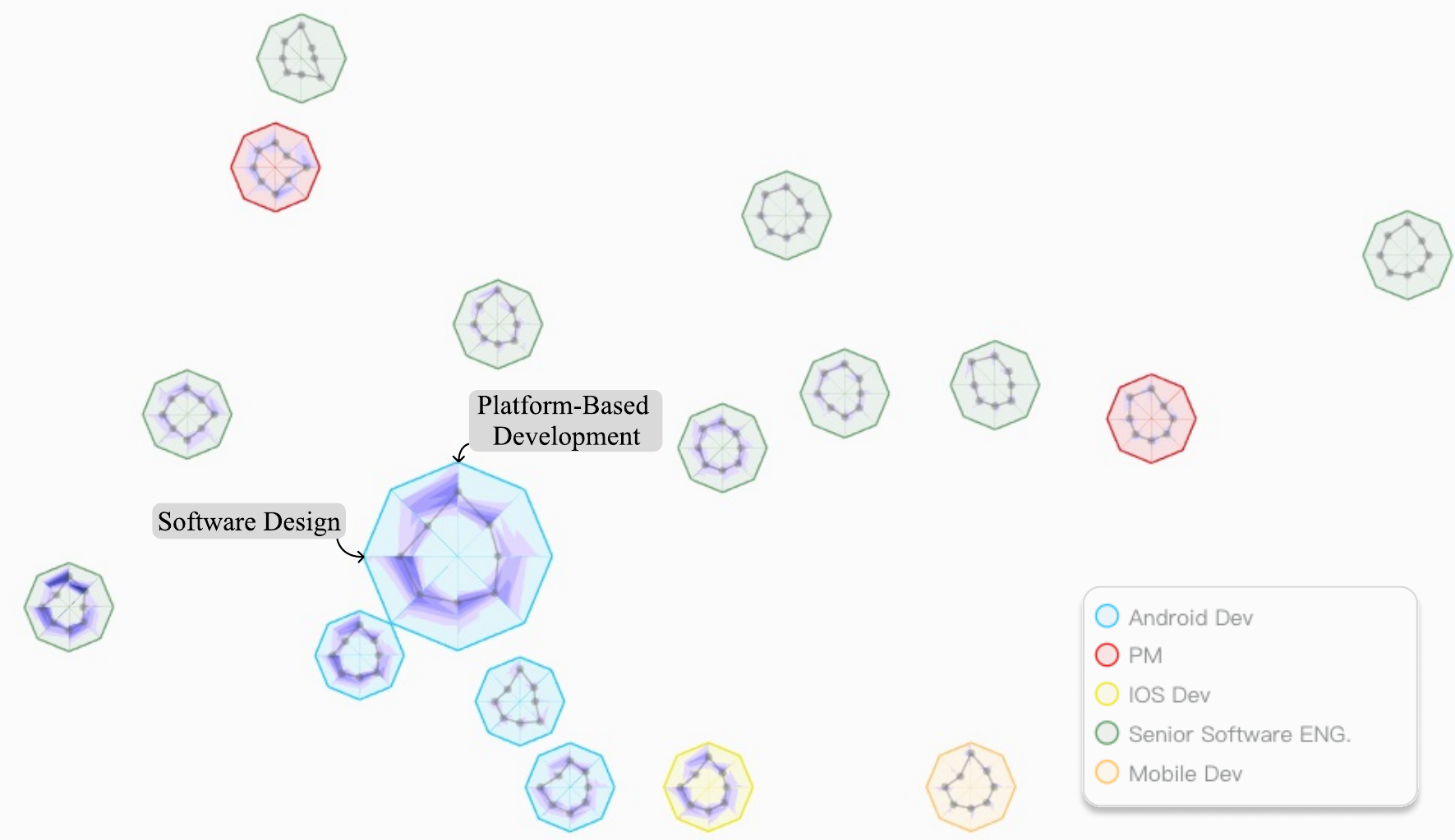}
    \caption{Inspecting and filtering job posts with skill patterns between job clusters.}
    \label{case2}
\end{figure}

\begin{figure}[ht]
    \centering
    \includegraphics[width=\linewidth]{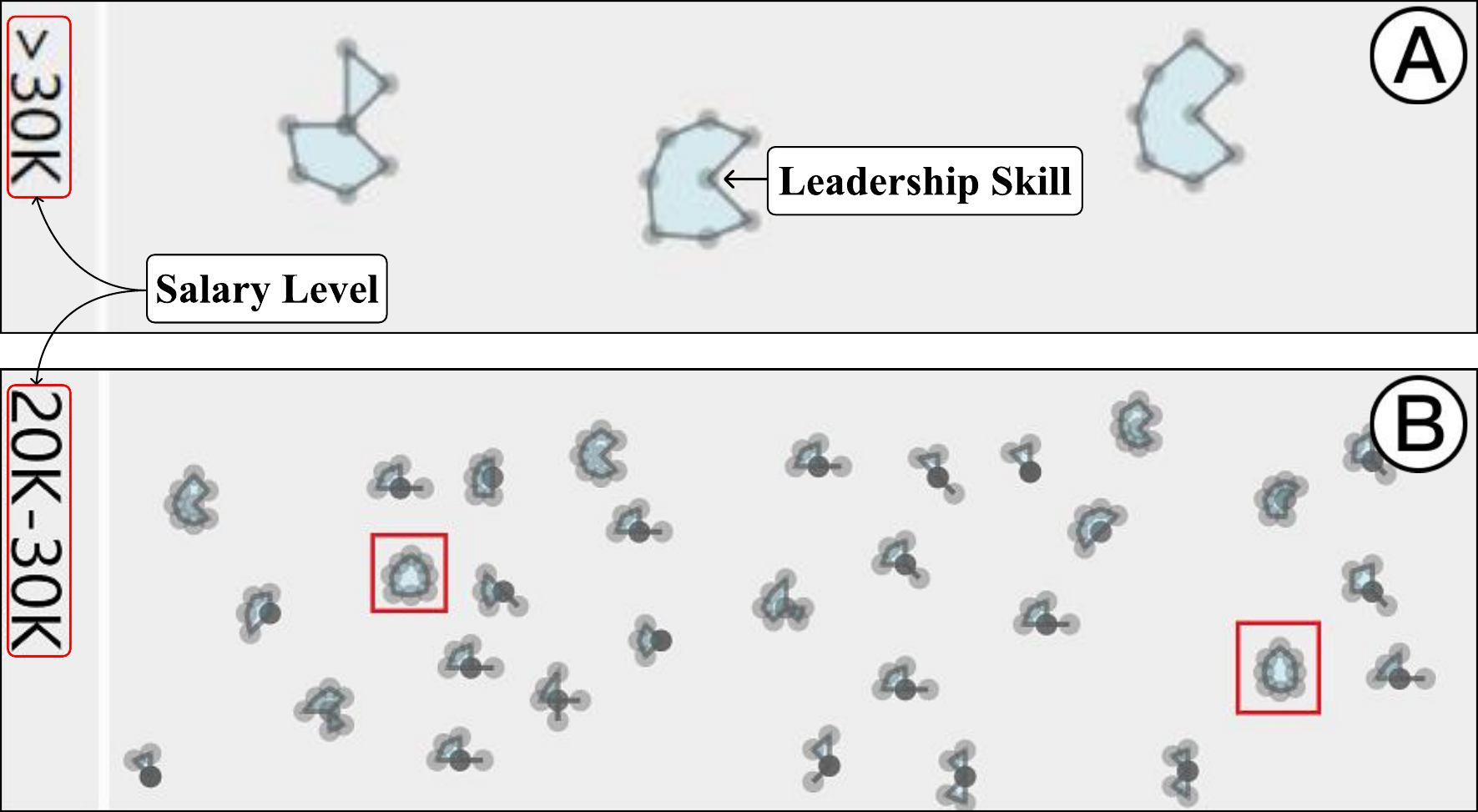}
    \caption{Inspection and comparing job posts with skill patterns between individual job posts.}
    \label{case2twoion}
\end{figure}

% \textbf{Filtering job posts mainly based on skill patterns and salary level.} 
\textbf{Filtering job posts by means of balancing the skill required and the salary provided.} 
In the Post Exploration View, as shown in Figure~\ref{case2}, P5 checked skill patterns of clusters with relatively dense job posts, which could provide a higher salary level according to his experience. Then he chose a cluster, namely, ``Android Dev''. It was found that ``Platform-Based Development'' and ``Software Design'' were highly required, which was a good news for an experienced job seeker. What's more, 
it also needed applicants with strong leadership, which meant it was a leadership position indeed. Therefore, by clicking the cluster, P5 inspected the job posts involved for more details, as shown in Figure~\ref{case2twoion}. In the Post Map, he found that most job posts with salary level higher than 30000 payment had a relatively low demand for the leadership capability, and thus he decided not to consider those job posts with attractive salary level. Afterwards, P5 found two job posts with high requirements for leadership capability in the salary range of 20000-30000, and compared them with left and right click respectively.

% \textbf{Comparing post details for the final decision-making.} 
\textbf{Prioritizing self-growth for his final decision of job application.} 
In the Post Detail View, it could be found that the two companies were located in Guangzhou and Shenzhen respectively, which were both P5 preferred. Then, P5 carefully compared the details the two job posts. Although the payment of the second job post was more attractive, the first one seemed to have more individual development space, and complete welfare and vacation system. Finally, P5 chose to send his resume to the first company.

Finally, P5 commented that the skill-driven system was very complete and easy to use with nice visual effect. It was surprising that calculations of skill demand and posts' similarities were quite accurate and reasonable. Compared with recruitment websites, this system emphasized on skills, helping job seekers find their satisfied job matched with his abilities efficiently.

%% file: 8-UserStudy.tex
\section{User Interviews}

To further evaluate the effectiveness of \toolName{}, we conducted in-depth user interviews with 26 target users who are hunting for a job.

\subsection{Study Design}

\textbf{Participants.} 
We invited 26 participants (12 females) from five different universities and four companies to join our in-depth user interviews. The participants are different from those who participated in the pilot study. Specifically, U1-12 and U13-20 are undergraduate and postgraduate students respectively majored in computer science and technology. U21-22 and U23-26 are from state-owned and private enterprises respectively, with more than at least one year working experience related to computer technology. To guarantee that the findings from the interviews are general for common users, none of the participants has a background in visualization or HCI. Some of the user interviews were conducted through the online tencent meetings due to the COVID-19 pandemic.

\begin{figure}[ht]
    \centering
    \includegraphics[width=\linewidth]{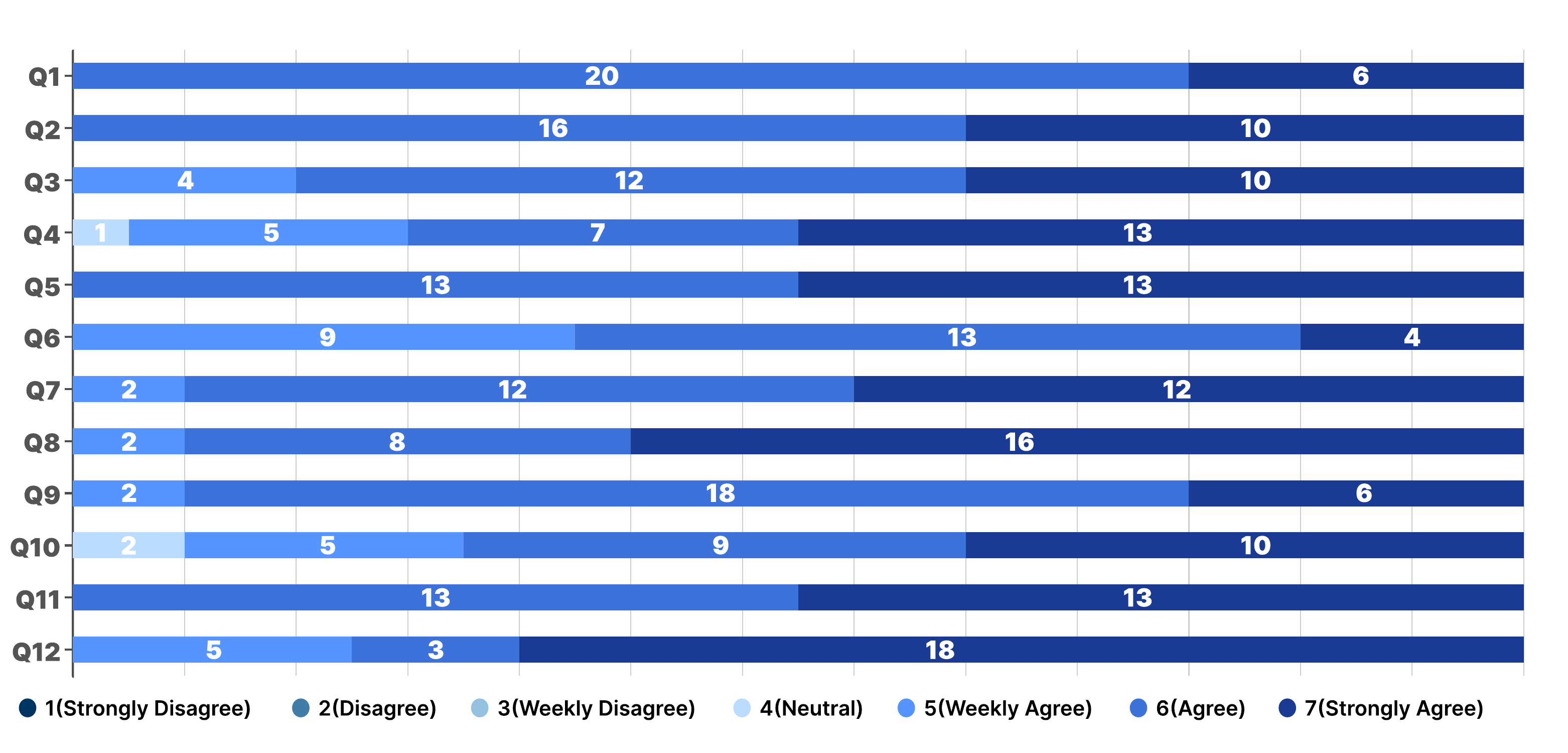}
    \caption{The statistic results for user interviews including effectiveness, visual design, interaction and usability.}
    \label{ustudy}
\end{figure}

\textbf{Tasks and procedures.}
We asked participants to fulfill five carefully-designed tasks to assess \toolName{}, as shown in Table 1. Each task corresponded to the use of each visual design (A1, A2, A3, B and C). The user interview for each participant was conducted using the online \toolName{} system. We recorded and took notes for each interview and their interaction processes. We first introduced the analysis workflow and the corresponding visual designs of \toolName{} to the participant. Then, we showcased an example to better illustrate the usage of \toolName{}. The above tutorial lasted for about 20 minutes. After that, the participants were asked to accomplish the pre-defined tasks in Table~\ref{table:1}. It worth noting that the tasks are essentially open-ended, and the participants accomplish the tasks according to their own skills. Upon the exploration, participants were encouraged to describe the reasons for the selection in a think-aloud manner. The aforementioned tasks lasted about 35 minutes, consisting of 10 minutes of free exploration, 20 minutes of task completion and 5 minutes of mutual communication. We also invited every participants to rate the \toolName{} system based on a 7-point Likert scale (1 ``strongly disagree'' to 7 ``strongly agree'') from four aspects shown in Table 2. Finally, we further conducted a post-study interview with each participants, which lasted about 30 minutes. The statistic results are shown in Figure~\ref{ustudy}.

\begin{table*}[htbp]
\centering
\caption{The tasks for participants to perform in our user interviews. All tasks are grouped by the analysis workflow.}
% \resizebox{\linewidth}{!}{
\begin{tabular}{l | l}\hline
T1 & Select candidate job posts based on types of skills. \\
T2 & Select candidate job posts based on skills similarity. \\
T3 & Filter candidate job posts based on the location, qualification, experience and industry.\\\hline
T4 & Filter candidate job posts according to skill patterns and post distributions of job clusters.\\
T5 & Compare candidate job posts according to their skills, salary and company category.\\\hline
\end{tabular}
\label{table:1}
\end{table*}

\begin{table*}[htbp]
\centering
\caption{The questionnaire consists of four parts: the effectiveness for \toolName{} (Q1-3), the visual design (Q4-6), the user interactions (Q7-8) and the usability (Q9-12).}
% \resizebox{\linewidth}{!}{%
\begin{tabular}{l | p{0.8\linewidth}}
\hline
Q1 & The system can facilitate the selection and filtering of job posts. \\
Q2 & The system enables the exploration and comparison of skill patterns and key properties of different job posts. \\
Q3 & The system provides details of the selected job post. \\
\hline
Q4 & The overall visual design is easy to understand. \\
Q5 & The hierarchical visual design is helpful for selecting candidate job posts. \\
Q6 & The design of the augmented radar-charts for skill patterns exploration is effective. \\
\hline
Q7 & The user interaction of the visualization is smooth. \\
Q8 & The user interaction of the visualization is intuitive and easy to use. \\
\hline
Q9 & The visual analytics system is easy to use. \\
Q10 & The visual analytics system is easy to learn. \\
Q11 & I would like to use the visual analytics system to hunt for a job in the future. \\
Q12 & I will recommend the visual analytics system to others who are looking for a job. \\
\hline
\end{tabular}%
% }
\label{table:2}
\end{table*}

\begin{figure}[ht]
    \centering
    \includegraphics[width=\linewidth]{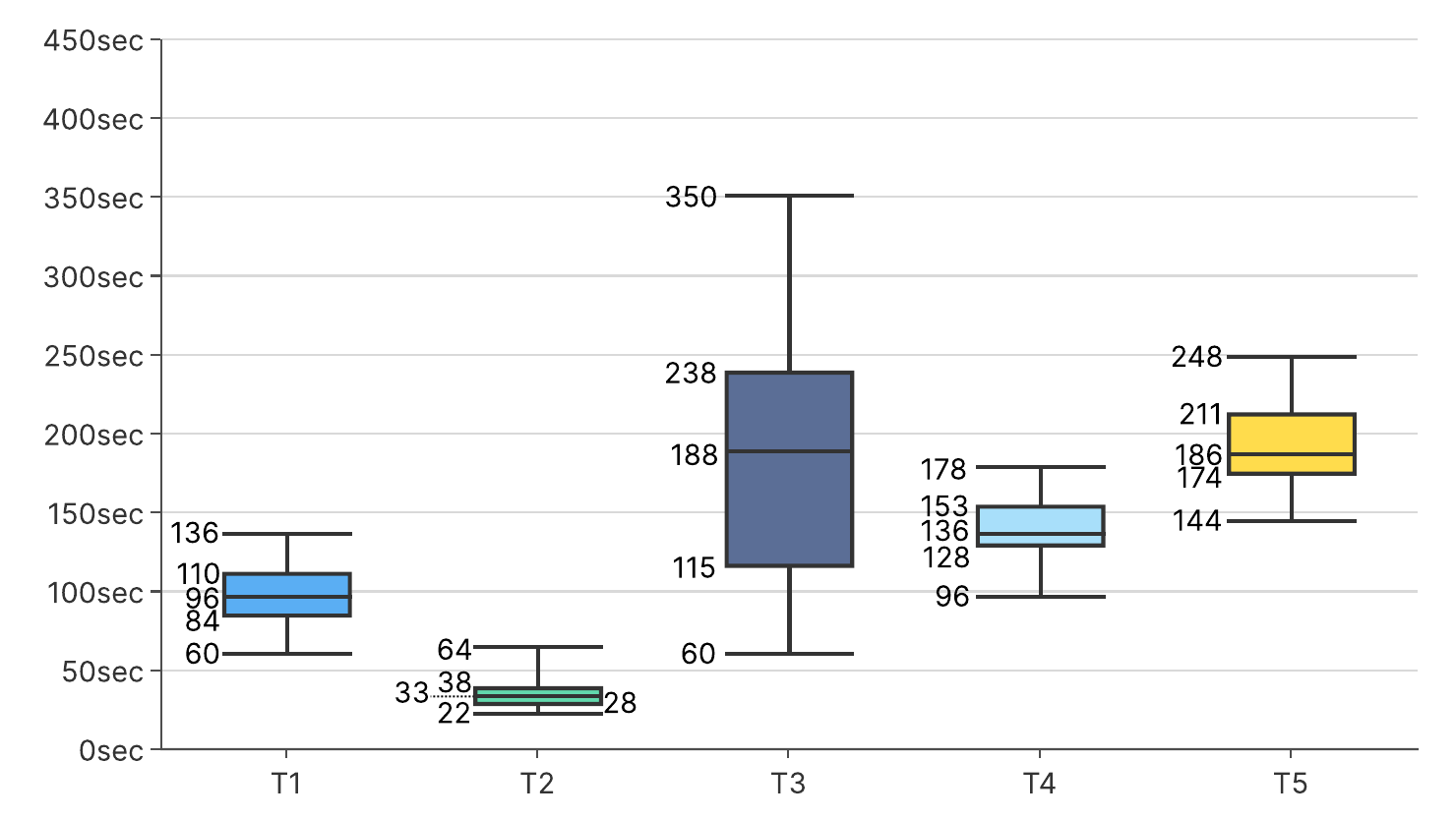}
    \caption{The time cost of different tasks.}
    \label{usertime}
\end{figure}

\subsection{Results}

We represent the description of each user task and completion time in Figure~\ref{usertime}. All participants completed any task within 17 minutes, while all tasks were finished in less than 200 seconds on average. Then we summarized all participants' detailed feedback as follows:

\textbf{Effectiveness for Exploration of Job Advertisements.}
Most participants agreed that \toolName{} can facilitate exploration of job advertisements($rating_{mean} = 6.28, rating_{sd} = 0.56$). In particular, all participants (U1-U26) highly appreciated the required skills analysis for job hunting. U19 commented, "\textit{It is beneficial that I can visually analyze skill needed of job posts for finding more opportunities. For example, by inspecting post similarities and skill patterns, I find the product manager position is also suitable for me except for Python developer, which is really intuitive.}" In addition, nine participants (U1-4, U6, U15-U18) commented that the proposed workflow is quite helpful. U6 said, "\textit{It is my first first time looking for a job. Due to lack of work experience, I really don't know which kinds of positions I am qualified for. Now \toolName{} enables me to seek a job starting from my skills instead of other external factors.}" U17 pointed out, "\textit{By exploring the job market with \toolName{}, I found that foundational skills are highly required by nearly all job posts. In particular, the skill of mathematics and statistics is emphasized by occupations of Web Developer, which I never thought of. And this was confirmed until communicated with an engineer with work experience of eight years.}"
U22 and U26 reported that \toolName{} can improve the efficiency of filtering job posts from a mass of job advertisements significantly compared with other online recruitment sites. Especially, \toolName{} facilitated inspecting the overall properties of job posts, and then comparing them according to their similarities and differences as a whole.

\textbf{Usability.}
Most participants 
% express a favorable disposition towards 
appreciate
the usability of \toolName{}($rating_{mean} = 6.30, rating_{sd} = 0.75$). U7-15 mentioned that \toolName{} is friendly and easy to use for university students. U19 also praised \toolName{}, "\textit{In my opinion, salary is not my first consideration, the suitable jobs are those best matches my skills and abilities, especially at the beginning of my career. To this end, I believe the interface is quite valuable and practical.}" Meanwhile, U21-22, who are preparing for job-hopping, commented, "\textit{\toolName{} will be helpful for planning my career development based on skills needed, because the dream job can be hardly offered instantly.}" U1 and U14 mentioned that they will try to submit resumes according to the positions selected with \toolName{}. Meanwhile, they expressed the desire to recommend \toolName{} to their classmates in the graduation season.

\textbf{Visual Design and Interactions.}
The majority of the participants appreciate the effective and user-friendly visual design, along with the flexible user interactions of \toolName{}($rating_{mean} = 6.29, rating_{sd} = 0.73$). Eight participants (U2, U8-13, U25) mentioned that the design of the augmented radar-chart is creative and interesting, which encodes skill patterns and distributions of different job posts simultaneously within limited space. Besides, all participants (U1-U26) agreed that the hierarchical visual design of Skill-job Overview is very informative. U10 commented, "\textit{Surprisingly, I can glanced at the view for overall situation of the job market, as well as skills needed and other major properties. It is very helpful for further job selection.}" For the user interactions, U1-26 confirmed that the system’s interactions are really smooth and easy to use.

\textbf{Suggestions.}
Despite the positive feedback, several participants also gave suggestions to improve \toolName{}. U7 and U24 suggested that the execution of job clustering is supposed to be more efficient. U20 pointed out the importance of constructing standard for skill evaluation. For example, what is a definition of mastering a programming language, and incorrect assessments of skills may lead to mismatches with jobs.

%% file: 9-Discussion.tex
\section{Discussion}

In this section, we summarize the lessons we learned
% during the development of \toolName{}. Then, we 
and
discuss the limitations of \toolName{}.

\subsection{Lessons}
% We learned many lessons from the system design and implementation.
During the development of \toolName{}, we mainly learned two lessons, which can be summarized as follows:

\textbf{Critical importance of visualization for skills needed.}
As shown in the above evaluations, \toolName{} received highly positive feedback from the participants. They emphasized the strong importance and potential of skill-driven visualization approaches for job hunting. As skill matching is a key factor for job hunting, the job seekers really need an interactive way to explore positions suitable for them in terms of skills effectively and efficiently. \toolName{} is only the first step to address such kinds of needs in both visualization and job hunting, which can be play a more significant role in the overall job market.

\textbf{``Less is More'' in visual designs matters much for job seekers.}
While designing the prototype of \toolName{}, we attempted to visualize all the properties of job posts simultaneously, which resulted in a sophisticated visual design. However, the three participants in Section~\ref{Requirements} pointed out that it is really confusing for them to explore the system because it may cost a lot of time to understanding the glyphs and how to interact with them for target users. Thus, we tried to strike a balance between the intuitiveness and expressiveness of visualization. We simplified the visual designs, and further proposed intuitive and novel designs, such as the tree-map design and the radar-chart to ensure that each target user can use it easily. Also, our designs are tailored to the target users. For example, in the skill-job overview, a skill framework highly relevant to the target users is employed to indicate the required competency for jobs in the market. The effectiveness of \toolName{} design is confirmed by the user interviews and their feedback. Most subjects found \toolName{} easy to learn and understand.

\subsection{Limitations}
Our evaluations have shown that \toolName{} can facilitate job hunting driven by skills effectively. However, the proposed approach still has limitations.

\textbf{Visual Scalability.}
According to the feedback of participants, target users will often select a small part of position types (e.g., up to 10) 
in the Skill-job Overview for most situations. Our case studies and user interviews have confirmed that \toolName{} can work well for these situations. However, the Post Exploration View of \toolName{} may suffer from scalability issues when it is used to explore job posts with an extreme number of position types,
% (e.g., 20 or even more), 
which can result in decreased clustering efficiency and reduced cluster quality. 
% which can impair the tool's responsiveness and the user's ability to derive meaningful insights.
It can affect the usability of \toolName{}. It's important to note that this limitation aligns with typical job-seeking behavior and doesn't significantly impact the tool's effectiveness for the vast majority of users. For scenarios requiring analysis of an extremely large number of job types, future work could explore advanced clustering techniques or alternative visualization methods to enhance scalability while maintaining usability.
% \wy{It is also necessasry to defend our design here. Pls refer to my response in the cover letter.}

\textbf{Generalizability.}
With standard skills framework in computer science and engineering, 
we take the job posts related to computer science and engineering as an example and used the job posts collected from 51Job in the evaluation of \toolName{}. 
% With standard skills framework in computer science and engineering, \toolName{} in our work is mainly tested using the job advertisements from the computer industry collected from 51Job.
However, it is important to be noted that 
\toolName{} has the potential to be extended to job posts of other industries (e.g., marketing, business and journalism), as the workflow and data dimensions of job advertisements are almost the same across different industries.
The major difference is that different fields have different skill framework, and it is necessary to construct the skill framework before applying \toolName{} to job posts of other industries. Furthermore, the augmented radar-chart glyph also demonstrates potential for broader applications beyond job market analysis, such as in financial portfolio management, healthcare patient monitoring, and product development, where quick comparison of multiple entities across various dimensions is crucial.

% In particular, since skills required by different majors and industries differ significantly, especially for professional skills, it is difficult to establish a skills framework in general. Thus, the key issue for extension of \toolName{} lays on the construction of the skill framework appropriate for the target major.

% \yong{I do not find a discussion on the learning curve of \toolName{}.}

%% file: 10-ConclusionAndFutureWork.tex
\section{Conclusion and Future Work}

We propose a skill-driven visual analytics approach, \toolName{}, to help job seekers efficiently explore the required skills and other relevant information of a large number of job posts. Instead of applying the filtering of basic job information on most job websites, our proposed approach can allow target users to find proper job advertisement posts matching their own skills. 
% In particular, a novel hierarchical visual design to visualize the skill sets, job posts, and the relationships between them, as well as an augmented radar-chart glyph to represent job posts in terms of both skills structure and posts distribution in a compact manner, are proposed to facilitate interactive exploration and analysis of job posts. 
Case studies and user interviews conducted by job seekers
% majored in computer science 
demonstrate the usefulness and effectiveness of \toolName{} in helping users gain deep insights into 
online recruitment information in a skill-centered manner.

In future work, we would like to extend \toolName{} to other languages and further evaluate its effectiveness with datasets of other languages. Also, our approach is mainly focusing on the computer-science related recruitment advertisements. It will be interesting to further extend our approach to other specialties.